\documentclass[sigconf]{acmart}

\AtBeginDocument{%
  \providecommand\BibTeX{{%
    \normalfont B\kern-0.5em{\scshape i\kern-0.25em b}\kern-0.8em\TeX}}}

\setcopyright{acmcopyright}
\copyrightyear{2018}
\acmYear{2018}
\acmDOI{XXXXXXX.XXXXXXX}

\acmConference[Conference acronym 'XX]{Make sure to enter the correct
  conference title from your rights confirmation emai}{June 03--05,
  2018}{Woodstock, NY}
\acmPrice{15.00}
\acmISBN{978-1-4503-XXXX-X/18/06}

\usepackage{xspace}

\usepackage{graphicx}
\usepackage{subfig}
\usepackage{adjustbox}
\usepackage{listings}
\usepackage{amsmath}
\usepackage{subcaption}
\usepackage{soul}
\usepackage{xcolor}
\usepackage{algorithm}
\usepackage{algpseudocode}
\usepackage{listings}
\usepackage{booktabs}
\usepackage{multirow}
\usepackage{enumitem}
\usepackage{bbm}
\definecolor{lightgray}{rgb}{0.95, 0.95, 0.96}
\lstdefinelanguage{SQL}%
  {morekeywords={%
      SELECT, FROM, WHERE, AND, INNER, JOIN, ON, AS, WITH},%
   sensitive=true,%
   morecomment=[l]--,%
   morestring=[b]',%
   morestring=[b]"}

\newcommand{\introparagraph}[1]{{\textbf{#1}}}

\begin{document}

\title{Hybrid Querying Over Relational Databases and Large
Language Models}




\author{Fuheng Zhao}
\affiliation{%
  \institution{UC Santa Barbara}
}
\email{fuheng\_zhao@ucsb.edu}
\author{Divyakant Agrawal}
\affiliation{%
  \institution{UC Santa Barbara}
}
\email{agrawal@cs.ucsb.edu}

\author{Amr El Abbadi}
\affiliation{%
  \institution{UC Santa Barbara}
}
\email{amr@cs.ucsb.edu}








\begin{abstract}
Database queries traditionally operate under the closed-world assumption, providing no answers to questions that require information beyond the data stored in the database. Hybrid querying using SQL offers an alternative by integrating relational databases with large language models (LLMs) to answer beyond-database questions. In this paper, we present the first cross-domain benchmark, SWAN, containing 120 beyond-database questions over four real-world databases. To leverage state-of-the-art language models in addressing these complex questions in SWAN, we present two solutions: one based on schema expansion  and the other based on user defined functions. We also discuss optimization opportunities and potential future directions. 
Our evaluation demonstrates that using GPT-4 Turbo with few-shot prompts, one can achieves up to 40.0\% in execution accuracy and 48.2\% in data factuality. These results highlights both the potential and challenges for hybrid querying. We believe that our work will inspire further research in creating more efficient and accurate data systems that seamlessly integrate relational databases and large language models to address beyond-database questions.

\end{abstract}

\ccsdesc[500]{Information systems~Query languages}
\ccsdesc[500]{Information systems~Information retrieval}
\ccsdesc[500]{Information systems~Information integration}
\keywords{Hybrid Query, Relational Databases and Large Language Models}




\maketitle

\section{Introduction}
\label{sec:intro}
The Relational model and SQL have achieved widespread acceptance and usage in data management systems. In particular, SQL continues to evolve by adding new syntax and features, enhancing its capabilities to meet the growing demands of modern data systems~\cite{Michael2024}. 
It is well known that database queries are evaluated under a closed domain assumption, meaning that the queries address aspects of the real world based solely on the data stored in the relational database management system~\cite{reiter1988should}.
While the direct approach is to say NO to beyond-database questions, researchers in our community have also investigated alternatives such as providing answers based on incomplete data with heuristic algorithms~\cite{levy1996obtaining} and crowdsourcing~\cite{franklin2011crowddb}. In this paper, we explore the use of large language models to address these questions.


In the natural language processing community, open-domain question answering has been extensively studied, often encompassing a broader range of inquiries. 
This long-standing task aims to provide factual answers to natural language questions by drawing from large, unstructured collections of texts and documents, such as Wikipedia. 
By pre-training on large corpus of knowledge, large language models (LLMs) have demonstrated significant potential in providing world knowledge and performing complex reasoning~\cite{hendrycks2020measuring, zhao2023llm}.
In this paper, we are specifically interested in beyond-database questions which have partial information grounded in the relational database. Unlike open-domain questions, beyond-database questions require integrating and reasoning with structured data from relational databases as well as drawing from large, unstructured collections of texts.

\begin{figure}[hbtp]
    \centering
    \includegraphics[width=\linewidth]{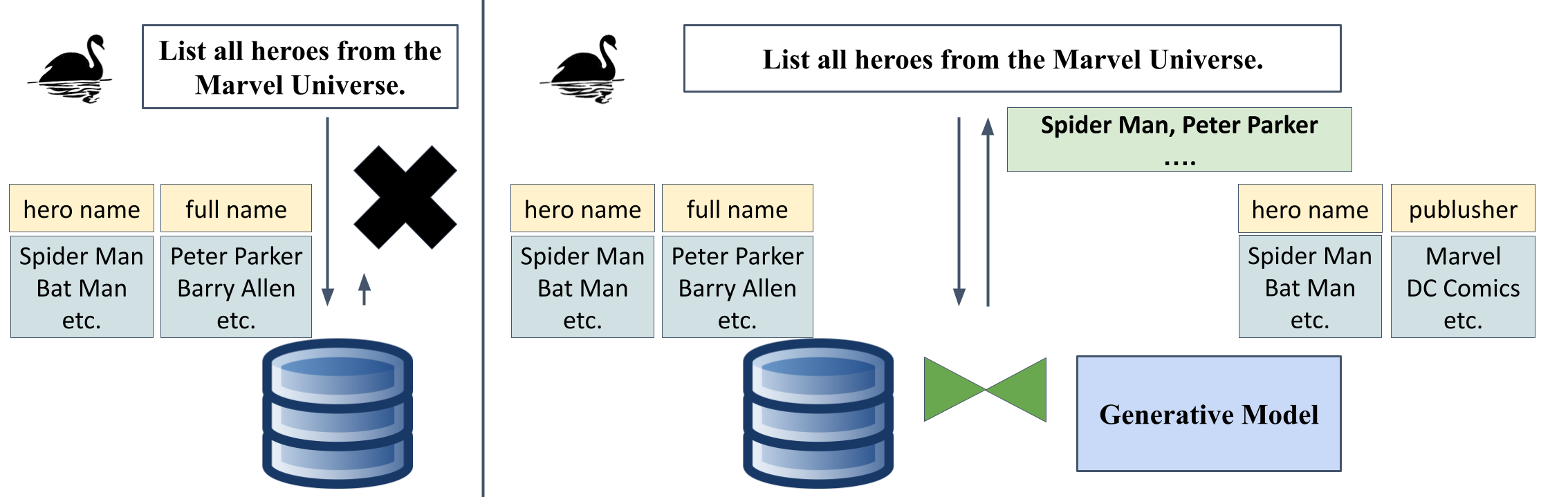}
    \caption{An illustrative example contrasting the answering of a beyond-database question solely using a database (left) versus hybrid querying over both databases and large language models (right).}
    \label{fig:motivationfigure}
\end{figure}

\textbf{A motivating example:} Consider a simple database with a single table containing superhero information, as shown in Figure~\ref{fig:motivationfigure}. The schema for the database is: \textit{superhero(hero\_name, full\_name)}. Suppose the user wants to list all the hero names from the Marvel Universe within the database. 
This would be considered as a beyond-database question, since the database contains relevant or partial information to the request but cannot directly provide the answer (i.e., which heroes are Marvel universe). 
On the other hand, large language models such as ChatGPT can be used to identify the publisher for each superhero characters. 
Assume we treat LLMs as a table containing the hero\_name and the publisher. Then, a SQL query like: "SELECT hero\_name, full\_name FROM LLM JOIN superhero ON LLM.hero\_name = superhero.hero\_name WHERE llm.publisher = 'Marvel';" 
Hence, hybrid querying by integrating relational databases and large language models offers a powerful approach for addressing beyond-database questions. 


In this work, we propose \textbf{SWAN}, \textbf{S}olving beyond-database queries \textbf{W}ith generative \textbf{A}I a\textbf{N}d relational databases, the first hybrid query benchmark. SWAN comprises 120 beyond-database questions and spans four big databases, covering four diverse domains. The original databases and questions are from the recent Bird benchmark\cite{li2024can}, which is a benchmark for evaluating natural language to SQL translation. The data in these databases are collected from open-source relational databases in platforms such as Kaggle. Also, the proposed SWAN benchmark challenges large language models to both select values from a given list (e.g., choose publisher name from a list of predefined publishers) and generate free-form outputs (e.g., determine the city based on a street address). 
In addition to proposing the SWAN benchmark, we also introduce HQDL, a preliminary solution for answering beyond-database questions. We evaluate HQDL over the state-of-the-art large language models such as ChatGPT (gpt-3.5-turbo)~\cite{ouyang2022training} and the GTP-4 turbo~\cite{opennew, achiam2023gpt}. Our experimental results indicate significant challenges for current state-of-the-art models in generating factual data and accurately answering beyond-database questions with hybrid queries. With in-context learning (ICL), GPT-4 Turbo achieves 40.0\% execution accuracy and 48.2\% of its generated data is factually correct.
Proposing a novel benchmark for beyond-database queries will focus attention and drive more research in this critical area.  It will also encourage researchers to develop innovative solutions that are founded on solid well understood foundations.
The evaluation results based on HQDL further highlight the need for more advancements in improving the accuracy and reliability of answering beyond-database questions with hybrid queries.

The paper is organized as follows: In Section~\ref{sec:background}, we discuss the background and related works on hybrid querying. Section~\ref{sec:swan} introduces the SWAN benchmark and provides details on the construction of the databases and the corresponding beyond-database questions. In Section~\ref{sec:approach}, we present HQDL a preliminary solution for utilizing large language models to solve these complex questions and discuss potential areas for improvement. Section~\ref{sec:eval} showcases our evaluation of HQDL on the SWAN benchmark. Finally, Section~\ref{sec:conclu} concludes the paper.

\section{Background and Problem Statement}
\label{sec:background}
In this section, we provide some basic background of large language models (LLMs) and introduce related work from both the database and the natural language communities for answering beyond-database questions.

\subsection{Large Language Models Basics}

Large language models are trained on vast datasets to produce high-quality responses based on input prompts. They have shown remarkable capabilities in addressing complex tasks across various domains, including logic reasoning~\cite{zhao2023llm}, natural language to SQL translations~\cite{pourreza2024din, gao2023text, Floratou2024NL2SQLIA}, and data system tuning~\cite{zhou2024db, giannakouris2024demonstrating}.

Large language models are frequently used to retrieve domain-specific knowledge and to simplify the information retrieval process by providing direct natural language answers. However, these models may exhibit hallucinations and  factuality errors. Consequently, recently many researchers have focused on enhancing the capabilities of LLMs to provide factual information~\cite{wang2023survey}. In-context learning (ICL) is a widely adopted strategy to enhance the factual accuracy of generated content. ICL enables a language model to learn from example demonstrations within its context to improve performance~\cite{brown2020language}.



\subsection{Related Work}
Answering beyond-database questions has been investigated in a variety of research efforts. CrowdDB~\cite{franklin2011crowddb}, Qurk~\cite{marcus2011crowdsourced}, Deco~\cite{park2012deco}, and hQuery~\cite{parameswaran2011answering} introduce crowdsourced query processing systems that address the closed-world assumption in traditional query processing. However, the cost of incorporating human input can be significant, impacting both time and resources. Inspired by the capabilities of LLMs, researchers have investigated whether LLMs can be used for declarative prompting~\cite{parameswaran2023revisiting} and data cleaning tasks such as data imputation where LLMs repair dirty or missing values in data entries~\cite{narayan2022can, chen2023seed}. While this is closely related to hybrid queries, hybrid querying over relational databases and LLMs presents its own set of challenges in combining structured and unstructured data sources, ensuring the correctness of generated data, and materializing the data for future uses.
Furthermore, two recent studies~\cite{urban2023omniscientdb, saeed2023querying} laid out the vision for augmenting relational databases with data generated from LLMs. However, due to the absence of an evaluation benchmark, both studies are limited to preliminary case studies. For instance, Galois~\cite{saeed2023querying} executed 46 SQL queries (drawn from the Spider benchmark~\cite{yu2018spider}) solely on LLMs, without involving any relational databases. Also, they manually verified the generation results, comparing the output table statistics (e.g., cardinality) and verifying content accuracy with the ground truth. Our proposed SWAN benchmark is built on top of databases collected in the Bird benchmark~\cite{li2024can}, which has 270x (on average) more rows compared to the databases in the Spider benchmark~\cite{yu2018spider}. Moreover, in addition to content accuracy, we also compare the execution accuracy among hybrid queries (see more explanations in Section~\ref{sec:eval})

\section{SWAN Construction}
\label{sec:swan}

\subsection{Databases in SWAN}
We constructed the SWAN benchmark\footnote{see https://github.com/ZhaoFuheng/SWAN/} based on the Bird benchmark~\cite{li2024can}. In the Bird benchmark, there are eleven diverse database domains. However, we have identified that many of these databases are overly narrow, and the questions are too specific to be answered effectively by a general intelligence model. For instance, the financial database includes detailed tables on bank accounts, credit card information, loans, and trading transactions. This level of specificity are out of the scope of a general AI's capabilities. As a result, we selected four diverse databases: European Football, Formula One, California Schools, and Superhero. These databases cover a broad range of topics, from sports statistics and history to educational trends and fictional characters.

\subsection{Schema Curation}
The main challenges in evaluating hybrid queries are: i) generating beyond-database questions, and ii) ensuring the availability of ground truth answers for these questions. Fortunately, the Bird benchmark provides valuable assets: natural language questions, SQL queries, and the databases. Leveraging these resources, we can modify the databases to create questions that the databases can not answer based on the database content. For the four selected databases, we removed specific columns or entire tables to generate beyond-database questions. For example, in the Superhero database, the $superhero$ table contains a $publisher\_id$ field, which is a foreign key used to identify the $publisher\_name$ in the $publisher$ table. By dropping the $publisher\_id$ column, all questions related to finding the name of the publisher become unanswerable based on the newly curated database. While the entire publisher table can be directly dropped, we kept the distinct values of $publisher\_name$ to assist LLMs in correctly formatting the output related to publishers (see more explanation in the next section). After schema curation, the statistics of the selected databases are shown in Table~\ref{tab:dbstats}. 


\begin{table}[hbtp]
    \centering
    \begin{tabular}{c|c|c|c}
        Database & Tables & Rows/Table & Cols Dropped \\
        \hline
         European Football & 7 &  31828 & 12\\
         Formula One & 13 & 39561 & 12 \\  
         California Schools & 3 &  9980 & 12\\  
         Superhero & 10 & 1061 & 11\\  
    \end{tabular}
    \caption{Statistics of databases in SWAN.}
    \label{tab:dbstats}
\end{table}

\subsection{Free Form Response and Value Selection}
In SWAN, the challenges for LLMs to generate factual data can be broken down into two categories: i) free form response and ii) value selection. The free form response requires LLMs to generate data when some context is provided. For instance, in the California Schools database, the tables originally contained both the school name and the school url. We removed the school url column, and as a result, we expect LLMs to generate short-form urls for the schools. Often, the school url is closely related to the school name and often ends with edu. Value selection involves choosing data values from a predefined list (e.g., a list of unique publisher names). For instance, after we removed the $publisher\_id$ field from the $superhero$ table and the entire $publisher$ table from the Superhero database, we retained the unique values of $publisher\_names$, which contains the names of all publishers for the superheroes in the database. Consequently, the list of all publisher names can be provided to the LLMs, allowing them to select the appropriate publisher for each superhero.

\subsection{Keys for Tables from LLMs}
\label{sec:keys}
In relational databases, a foreign key column is often represented as an integer linked to the primary key column in another table. However, integers do not provide any meaningful insights for LLMs to generate useful data values. According to SQL standards, a foreign key must reference a unique key in the foreign table. Therefore, we have curated the databases to include meaningful foreign keys for the data generated by LLMs. For example, in the Superhero database, we assume the combination of $superhero\_name$ and $full\_name$ of a superhero serves as the key to finding the publisher information. Also, we have ensured that there are no duplicate pairs of ($superhero\_name$, $full\_name$) in the table. Our approach of designing meaningful keys for LLMs to generate data aligns with the data model in crowd-sourcing systems. For example, Deco's Fetch/Resolution rules~\cite{park2012deco} use meaningful keys as input and ask crowd-source workers to generate a group of attributes based on the given keys.

\subsection{Beyond-Database Questions}
For each database in SWAN, we provide 30 beyond-database questions, resulting in a total of 120 questions across all databases. For each question, we also supply i) a hybrid SQL query that joins the tables in the relational database with the tables generated by LLMs (assume the values generated by LLMs are materialized as tables), ii) a hybrid SQL query that directly invokes LLM calls based on BlendSQL~\cite{glenn2024blendsql} functions, and iii) the corresponding gold SQL query from Bird, such that the expected answer is the execution results of the gold SQL query on the original Bird databases. 
Notably, these hybrid queries are manually crafted and fully executable. They are provided to assess the current capabilities of combining LLMs with databases to answer beyond-database questions. Automating the translation of beyond-database questions into hybrid queries is left as future work.
\section{Answering Beyond-Database Questions}
\label{sec:approach}
In this section, we discuss two different approaches to answer beyond-database questions in SWAN. One is based on schema expansion and the other is based on SQL user defined functions. At the end of this section, we discuss the promising optimization opportunities of these solutions and outline potential directions for future research.

\subsection{HQDL}
First, we introduce Hybrid Query Database and LLM (HQDL), a preliminary solution for solving beyond-database questions based on schema expansion. Given a beyond-database $NL$ question, one can expand the database schema by including new columns or new tables such that $NL$ question becomes answerable based on the new schema. Then, LLMs can be used to fill in all the missing data entries after schema expansion. Based on the newly updated schema, one can write a regular SQL query to answer question $NL$ directly.

\subsubsection{Data Generation}
For each database, SWAN provides a list of missing columns and tables that need to be generated by LLMs. 
SWAN has also provided the keys consisting of the minimal number of attributes that represent the primary-key/foreign-key (PK-FK) relationships between the existing tables in the relational database and the tables generated by LLMs. This ensures that the necessary keys can be used as input for the LLMs, enabling them to accurately generate the missing data entries.
We would also like to note that these information can be helpful, though its use is optional. In HQDL, we choose to directly use these metadata information. 
An example of a zero-shot prompt used to generate missing data values in the Super Hero database is provided below. This prompt instructs LLMs to infer and fill in missing data entries by supplying guidance, column names, and example values for certain columns (e.g., publishers and colors).

\begin{lstlisting}[language=SQL]
Your task is to fill in the missing values in the target entry from the `superhero` database.
Return a single row with no explanation.

The columns are: `superhero_name`,`full_name`,`eye_color`,`hair_color`,`skin_color`,`publisher_name`,`race`,`gender`,`moral_alignment`,`powers`

The possible values for `publisher_name` are[Dark Horse Comics', 'DC Comics', Marvel Comics', ...]

Value list of colors, power names, etc.

Target Entry:'{superhero_name}','{full_name}',?,?,?,?,?,?,?,?
The output should consist of a single row containing 10 fields.
Answer:
\end{lstlisting}

HQDL needs to instruct the LLMs to fill in the missing values in the target data entry. HQDL also adopts the widely accepted 'No Explanation' rule introduced by OpenAI~\cite{OpenAIPlatform2024}, which consistently improves the quality of generated answers for semantic parsing~\cite{gao2023text}. Furthermore, HQDL provides value lists, such as publishers and colors, for LLMs to select from. Since only the id fields are removed (e.g., $publisher\_id$), HQDL can directly retrieve all these predefined data values. The goal is to avoid ambiguous data values such as 'Marvel' v.s. 'Marvel Comics' in which both values represent the same publisher but pose challenges for automatic evaluation. 

In addition to zero-shot prompts, we also conduct investigations on few-shot prompts. A one-shot prompt for generating the missing data values for the Super Hero database is provided below:
\begin{lstlisting}[language=SQL]
Prefix (instructions and value lists)

/*An example is provided before the target data entry*/
Example Entry:'3-D Man','Charles Chandler',?,?,?,?,?,?,?,?
Example Answer:'3-D Man','Charles Chandler','Brown','Grey','No Colour','Marvel Comics','-','Male','Good','Agility,Super Strength,Stamina,Super Speed'

Target Entry:'{superhero_name}','{full_name}',?,?,?,?,?,?,?,?

Answer:
\end{lstlisting}
As shown above, an example data entry and the corresponding answer are provided to the LLM for the constructed record corresponding to '3-D Man' and Charles Chandler. In the evaluation section (Section~\ref{sec:eval}), we will show that few-shot demonstrations significantly improve the quality of the generated data entries."

{\it Data Extraction.}
After collecting all data entries generated by the LLMs, HQDL materializes these entries into tables. HQDL uses the Python csv module's reader to process these entries, converting them into a structured format, and inserting them into new tables in the underlying SQLite database. Moreover, in SWAN, there are both one-to-one and one-to-many relationships. When one-to-many relationships occur, HQDL condenses the tuples in the "many" side of the relationship into a long text. For example, each superhero may be associated with many powers. HQDL would condense all the powers into a long string separated by commas (e.g., "Agility, Super Strength, Super Speed").

\subsection{Hybrid Query UDFs}
We observe that industry has started integrating LLM calls directly into SQL syntax through user defined functions, such as DucksDB~\cite{raasveldt2019duckdb} and Google BigQuery~\cite{kashyap2023machine}. For instance, finding all hero names from the Marvel universe within the database can be rewritten as follows in Google BigQuery:
\begin{lstlisting}[language=SQL]
SELECT T1.full_name, T1.hero_name
FROM superhero AS T1 
JOIN ( SELECT publisher
    FROM ML.GENERATE_TEXT(
        MODEL `a_generative_model`,
        (SELECT CONCAT(prompt, superhero.hero_name) AS input_text
          FROM superhero ),
        STRUCT(0 AS temperature)
      ) ) AS T2 ON T1.hero_name = T2.publisher
WHERE
  T2.publisher = 'Marvel';
\end{lstlisting}
Hybrid querying through UDFs offer more control for the database to optimize the hybrid query, build materialized views, and potentially reduce the amount of data generated by LLMs.

Since all four databases utilize SQLite, we can directly leverage BlendSQL~\cite{park2023generative}, an extended version of the SQLite relational database management system that supports LLM functions. In SWAN, we provide 120 hybrid queries using the BlendSQL syntax, enabling SWAN to evaluate current systems in querying both relational databases and LLMs.

\begin{table*}[htbp]
\centering
\caption{HQDL Execution Accuracy results on the SWAN benchmark using different number of demonstrations. The numbers in brackets report the accuracy improvement compared to the zero shot method.}
\label{hqdlEX}
\begin{tabular}{@{}cccccccc@{}}
\toprule
Model & Demonstrations &  California Schools & Super Hero &  Formula One & European Football & Overall \\ \midrule
\multirow{4}{*}{GPT-3.5 Turbo}  & 0-shot &  50.0\%  & 13.3\%     &  16.7\%    &  16.7\%    & \textbf{24.2\%}   \\

& 1-shot &  50.0\%  & 23.3\%     &  46.7\%    &  26.7\%    & \textbf{36.7\%} (+12.5\%)    \\

 &  3-shot   &  46.7\%  & 20.0\%  & 46.7\%  & 33.3\% & \textbf{36.7\%}  (+12.5\%)   \\

 &  5-shot   &  53.3\%  & 20.0\% &  46.7\%    & 33.3\%  & \textbf{38.3\%}  (+14.1\%)   \\  

 \midrule

 \multirow{4}{*}{GPT-4 Turbo}  & 0-shot &  50.0\%  & 23.3\%     &  36.7\%    &  16.7\%    & \textbf{31.6\%}  \\

 & 1-shot &  43.3\%  & 23.3\%     &  50.0\%    &  23.3\%    & \textbf{35.0\%} (+3.3\%)    \\

 &  3-shot   &  50.0\%  & 26.7\%  & 50.0\%  & 26.7\% & \textbf{38.3\%}  (+6.7\%)   \\

 &  5-shot   &  56.7\%  & 23.3\% &  50.0\%    & 30.0\%  & \textbf{40.0\%}  (+8.4\%)   \\

\bottomrule
\end{tabular}
\end{table*}

\subsection{Optimization Opportunities}
\label{sec:limit}
While these two solutions (HQDL and Hybrid Query UDFs) can be used to solve the challenges presented in our proposed SWAN benchmark, we believe that there are opportunities to improve upon these two solutions. 

First, to answer these beyond database questions, what contexts, other than the necessary keys and the predefined value lists, should be presented in the prompt to reduce LLMs hallucination? 
There are other attributes inside the relational database that may be relevant and it remains an open question on how to select the best context. One possible approach is to build a vector index on the database values or rows and then fetch the relevant information based on embedding similarity~\cite{zhang2023large, edge2024local}. 
Second, the prompts and static examples used in HQDL and Hybrid Query UDFs are hand-crafted. It would be more convenient for users if the data system may automatically generate prompts and examples based on the specific context and query requirements. A promising direction is to develop a principled declarative prompt engineering
toolkit~\cite{parameswaran2023revisiting}.
HQDL requires LLMs to generate and materialize all missing data, while Hybrid Query UDFs, through BlendSQL, optimize queries by pushing down predicates to avoid generating unnecessary data entries. Additionally, reusing previously generated data in HQDL is straightforward. In BlendSQL, generated data is cached by mapping the LLM input prompt to its output data. However, prompts with similar meanings (e.g., "Is the superhero from the Marvel Universe?" versus "Does the hero come from Marvel?") cannot directly reuse previous results. A promising approach to address this is incorporating query rewriting within Hybrid Query UDFs to fully leverage all cached LLM-generated data~\cite{zhao2023llm}. Query optimization and caching are essential for reducing costs and increasing throughput, making hybrid queries more accessible and efficient. BlendSQL currently implemented batching—retrieving data values for multiple rows in a single LLM call—and plans to support parallelized LLM calls in the future to further minimize query latency.


\section{Evaluation}
\label{sec:eval}





\begin{table*}[htbp]
\centering
\caption{HQ UDFs evaluation results on the SWAN benchmark. The numbers in brackets report the accuracy improvement compared to the zero shot method.}
\begin{tabular}{@{}cccccccc@{}}
\toprule
Model & Demonstrations &  California Schools & Super Hero &  Formula One & European Football & Overall \\ \midrule
\multirow{2}{*}{GPT-3.5 Turbo}  & 0-shot &  10.0\%  & 23.3\%     &  30.0\%    &  10.0\%    & \textbf{18.3\%}   \\

 &  5-shot   &  13.3\%  & 23.3\% &  43.3\%    & 3.3\%  & \textbf{20.8\%}  (+2.5\%)   \\  

\bottomrule
\end{tabular}
\label{hqUDFEX}
\end{table*}











\subsection{Evaluation Metrics}
In the context of evaluating hybrid queries, we propose three metrics: execution accuracy (EX), data factuality, and the number of input/output tokens used by the LLMs.

\introparagraph{Execution Accuracy (EX).} EX is a well accepted metric in the domain of semantic parsing~\cite{Floratou2024NL2SQLIA}. EX measures the percentage of hybrid queries that produce identical results to the ground truth (execution results from the Gold, correct, SQL). Since producing identical results is the end goal of hybrid querying, we adopt the EX metric.


\begin{table}[htbp]
\centering
\caption{The average F1 score for measuring the factuality of the generated data using HQDL.}
\begin{tabular}{@{}cccccccc@{}}
\toprule
Model & Demonstrations &  Average \\ \midrule
\multirow{4}{*}{GPT-3.5 Turbo}  & 0-shot  & \textbf{20.9\%} \\

 &  1-shot  & \textbf{37.3\%}    \\

 &  3-shot  & \textbf{41.4\%}    \\

 &  5-shot  & \textbf{42.7\%}  \\  
 \midrule

 \multirow{3}{*}{GPT-4 Turbo}  & 0-shot &  \textbf{29.3\%}    \\

 &  1-shot  &  \textbf{47.0\%}     \\

 &  3-shot   & \textbf{47.1\%}   \\

 &  5-shot   & \textbf{48.2\%}  \\  

\bottomrule
\end{tabular}
\label{tab:factual}
\end{table}

\introparagraph{Data Factuality.} We use exact string match to verify the data factuality for each data cell value. Because of the one-to-many relationships (the key from a table maps to many values generated by LLMs), we use the widely accepted F1 score, which is a harmonic mean of precision and recall, to measure the overall factuality of generate data entries for each database~\cite{wang2023factcheck}.

\introparagraph{Input and Output Tokens.}  We report the number of input and output tokens (i.e., words, sub-words) used in HQDL and Hybrid Query UDFs, which determine the monetary cost. For instance, GPT 3.5 Turbo priced at \$3 per million input tokens and \$6 per million output tokens.

\subsection{Experiment Configurations}

We evaluate HQDL and Hybrid Query UDFs on several OpenAI models (i.e., GPT-3.5 Turbo and GPT-4 Turbo) via OpenAI api calls. In all requests, we set the temperature to 0.

\introparagraph{Few Shots Demonstrations.} 
In the few-shot prompts, we provide static examples randomly selected from the original database. For HQDL, the few shots demonstrations are organized as static rows. In Hybrid Query UDFs, the few shots demonstrations are organized as a natural language question, an example database key, and the answer to the natural language question on the example database key (e.g., question: What is the driver code, key: Lewis Hamilton, and answer: HAM).

\subsection{HQDL Results}
In this subsection, we report and analyze the EX scores and the generated data factuality using F1 score.

\introparagraph{Zero Shot.}
As shown in Table~\ref{hqdlEX}, in terms of overall accuracy over all databases using 0-Shot demonstrations, GPT-4 Turbo achieves 31.6\% accuracy on the proposed SWAN benchmark, surpassing GPT-3.5 Turbo by 7.4\%. One major challenge in using zero-shot prompts to generate data entries lies in ensuring that the output format is consistent, as this significantly impacts the ease of data extraction. Despite specifying the number of fields need to be returned in the input prompt, LLMs sometimes return too few or too many fields and may occasionally return an empty string for a field. 
Moreover, we can observe from Table~\ref{hqdlEX} that for the California Schools database, GPT-4 Turbo achieved the same execution accuracy as GPT-3.5 Turbo. Questions in the California Schools frequently ask for the top schools. Consequently, queries answering these questions often include a LIMIT clause to retrieve only the top results. 
Hence, generating more accurate content for irrelevant entries does not necessarily lead to improvements in query execution accuracy. This observation motivated us to further examine the data factuality using F1 scores.

\introparagraph{Few Shot.}
The distinction between the few-shots and zero-shot experiments is the inclusion of several static examples. We know that the capabilities of language models can be increased with examples provided for in-context learning~\cite{brown2020language}. Hence, we expect LLMs to generate more accurate data with few shots, and the execution accuracy should improve as more examples are provided. As shown in Table~\ref{hqdlEX}, in general the execution accuracy improves for both models as more examples are provided. When the prompt contains 5 static examples, GPT-3.5 Turbo achieves 38.3\% and GPT-4 Turbo achieves 40.0\% execution accuracy, 14.1\% and 8.4\% accuracy improvements compared to the zero shot method. 

It is also interesting to note that both models achieve the highest execution accuracy on the California Schools and also the lowest execution accuracy on Super Hero database.
One-third of the queries in California Schools database contain a LIMIT clause, retrieving the top schools. In contrast, many questions in the Super Hero database seek specific superheroes (e.g., heroes from Marvel or those with blue eyes), and only about one-tenth of the queries for this database include a LIMIT clause. One explanation for execution accuracy difference between questions in California Schools databse and questions in Super Hero database is that LLMs may exhibit biases, as previous research has shown that they tend to favor higher socioeconomic entities~\cite{manvi2024large}. 
For instance, while LLMs can accurately identify schools with the highest standardized testing scores, they may struggle to identify schools with average or below-average grades. Because many queries in the California Schools database contain a LIMIT clause, even when an LLM provides inaccurate answers for many schools, the top results may still appear correct, masking potential errors in the model's full response.



\introparagraph{Data Factuality}
To measure data factuality (using the F1 score), we use exact string matching to compare the generated data with the ground truth for each cell. Also, we compute the average F1 score over all cells for each database. When the generated content is identical to the ground truth, then it scores a 100\% F1 score.
As shown in Table~\ref{tab:factual}, GPT-4 Turbo consistently generates more factual information than GPT-3.5 Turbo using the same prompt. For instance, with the 5-shot prompt, GPT-4 Turbo scores 5.5\% higher than GPT-3.5 Turbo. Also, the results clearly showcase that providing more examples in the input prompt increases the factuality of the generated output, which also leads to higher execution accuracy when executing the hybrid queries.

\subsection{Hybrid Query UDFs Results}
We evaluated the performance of BlendSQL~\cite{park2023generative} on the SWAN dataset to assess its effectiveness with hybrid query UDFs. Notably, on GPT-3.5 Turbo, the execution accuracy for 0-shot and 5-shot settings reached 18.3\% and 20.8\% (see Table~\ref{hqUDFEX}), which are lower compared to HQDL’s results of 24.2\% and 38.3\%. In our evaluation of hybrid query UDFs, we provided the keys and instructed the LLM to predict only the necessary information (most of the time a single cell value). This approach contrasts significantly with HQDL, where the LLM is given the key and tasked with predicting all column values for the corresponding row. Predicting all column values may be more advantageous than predicting a single column value, as it mirrors a chain-of-thought process that enables the model to leverage inter-dependencies between columns, thereby enhancing accuracy and coherence in its predictions. In HQDL, each LLM call generates a single row. In contrast, BlendSQL uses a default batch size of 5, where each request combines five keys into a list, prompting the LLM to return a list of five data entries corresponding to the five keys. Although batching reduce the number of LLM calls, it also increases the potential for errors, as processing multiple entries in a single call may lead to inaccuracies in the returned data~\cite{cheng2023batch}.

Another noteworthy difference between HQDL and HQ UDFS is in the format of the few-shot examples. In HQDL, we included static rows in the prompt as few-shot demonstrations, and the goal is to showcase the model how to complete a row of data based on the given keys. In contrast, for HQ UDFS (e.g., BlendSQL), we curated a list of question-answer pairs for each databases, and then BlendSQL selects relevant examples based on similarity metrics (e.g., cosine similarity using a sentence transformer) to find the most similar questions. For example, a demonstration question in the HQ UDFS system might be: 'Provide the city name based on the address.' Along with this question, given the address '5328 Brann Street', the expected city name is 'Oakland'.

\subsection{Evaluation Costs}
The monetary costs and the system's performance (e.g., latency and throughout) are implicitly determined by the number of input and output tokens. Here, we report the total number of input and output tokens for HQDL and HQ UDFs.

\begin{table}[hbtp]
    \caption{Total tokens used for HQDL and HQ UDFs for zero shot experiments.}
    \centering
    \begin{tabular}{c|c|c}
        \toprule
         Algorithm & Input Tokens & Output Tokens  \\
         \hline
         HQDL & 6.3 M & 1.5 M \\
         \hline
         HQ UDFs & 23 M & 2 M \\
         \bottomrule
         
    \end{tabular}
    \label{tab:tokens}
\end{table}

HQDL generates data entries for all the missing columns. As shown in Table~\ref{tab:tokens}, using zero-shot prompt, a total of 6.3 million tokens are used as inputs for LLMs, and 1.5 million tokens are generated by LLMs. On average, about 52k input tokens and 12k output tokens are used per beyond-database questions. If the number of beyond-database questions increases, the cost per question will decrease. 

For HQ UDFS, we expected it to use less tokens compared to HQDL, because HQ UDFS give more control to the database query optimizer. This allows the system to intelligently minimize token usage by pushing down predicates, meaning it generates tokens only for the specific data cells needed to answer the query. Surprisingly, for zero-shot prompt, the total input and output tokens are 23 million and 2 million respectively, as shown in Table~\ref{tab:tokens}. Compared to HQDL, HQ UDFs uses 3.6x more input tokens and 1.3x more output tokens.

The increased costs of HQ UDFs can be attributed to its limited use of cached results. For instance, to answer the beyond-database question, 'What is the height of the tallest player?', HQ UDFS used the LLMs to generate heights for all players, as the database lacks this information. Later, another question asks, 'Please list player names who are taller than 180cm.' The corresponding hybrid query created for this question prompts the LLMs to answer 'Is the player taller than 180cm?' However, it is evident that the previously generated heights could be directly reused to answer this question, rather than generating new responses. In HQ UDFS, LLM-generated content is cached as a mapping from input prompts to LLM output answers, making it challenging for the system to efficiently reuse cached outputs. In contrast, HQDL stores LLM-generated outputs directly as entities within relationships (schema expansion), simplifying reuse for users.

\section{Conclusion and Future Work}
\label{sec:conclu}

In this paper, we present the first benchmark, \textbf{SWAN}, for evaluating hybrid queries that answer beyond-database questions using relational databases and large language models. In addition to the benchmark, we also introduce HQDL, a preliminary solution for answering questions in SWAN based on schema expansion. We also provide queries to evaluate current Hybrid Query UDFs systems (e.g., BlendSQL). Our evaluation demonstrates that there are still many opportunities for improving the execution accuracy and also increasing the overall efficiency. 
To improve the execution accuracy and ensure high data fidelity, retrieve-augmented generation (RAG)~\cite{li2024enhancing} and supervised fine-tuning~\cite{elaraby2023halo, razumovskaia2024dial} are two promising direction to be integrated in hybrid querying systems. Moreover, there are numerous opportunities to optimize the pipeline for executing hybrid queries, increasing throughput and lowering monetary costs. For example, to further reduce costs and improve system throughput, it is essential to focus on: (i) implementing asynchronous and parallel hybrid query execution, (ii) designing improved caching mechanisms, and (iii) fully utilizing cached content. 

Additionally, in the current benchmark, we provided the missing columns for schema expansion and pre-written queries for both HQDL and HQ UDFS.
In future work, the process of answering beyond-database questions should be fully automated. Given a natural language question, LLMs should first evaluate whether it can be answered using the existing schema. For questions requiring information beyond the current database, LLMs could be designed to automatically expand the schema, populate missing values, and generate a SQL query (similar to HQDL) or construct a SQL query with user-defined functions to directly prompt LLMs for required information in real time.

We envision that our benchmark, the two baseline solutions, and the discussions on future optimization opportunities will spark interests within the community to develop comprehensive data systems that leverage the full potential of relational databases and large language models.

\begin{acks}
We thank the anonymous reviewers for their valuable feedback and Parker Glenn for assistance with running BlendSQL on SWAN. Fuheng Zhao was partially funded by Microsoft PhD Fellowship.
\end{acks}

\bibliographystyle{ACM-Reference-Format}
\bibliography{ref}


\end{document}